\documentclass{emulateapj}

\newcommand{\paperacknowledge}{We thank the referee for a helpful report. We are grateful to Roman Rafikov and Kaitlin Kratter for helpful discussions. JEO acknowledges support by NASA through Hubble Fellowship grant HST-HF2-51346.001-A awarded by the Space Telescope Science Institute, which is operated by the Association of Universities for Research in Astronomy, Inc., for NASA, under contract NAS 5-26555.}

\usepackage{color}

\newcommand{\bc}{}

\shorttitle{Disk-fed giant planet formation}
\shortauthors{Owen, J. E. \& Menou, K.}

\begin{document}


\title{Disk-Fed Giant Planet Formation}


\author{James E. Owen\altaffilmark{1}}
\affil{Institute for Advanced Study, Einstein Drive, Princeton, NJ 08540, USA}
\email{jowen@ias.edu}
\and
\author{Kristen Menou}
\affil{Centre for Planetary Sciences,
    Department of Physical \& Environmental Sciences, University of
    Toronto at Scarborough, Toronto, Ontario M1C 1A4,
    CANADA\\
    Department of Astronomy \& Astrophysics,
    University of Toronto, Toronto, Ontario M5S 3H4, CANADA}
\email{kristen.menou@utoronto.ca}
%
%


\altaffiltext{1}{Hubble Fellow}


\begin{abstract}
Massive giant planets, such as the ones being discovered by direct
imaging surveys, likely experience the majority of their growth through a
circumplanetary disc. We argue that the entropy of accreted material
is determined by boundary layer processes, unlike the ``cold-'' or
``hot-start'' hypotheses usually invoked in the core accretion and direct
collapse scenarios. A simple planetary evolution model illustrates how
a wide range of radius and luminosity tracks become possible,
depending on details of the accretion process. Specifically, the proto-planet evolves towards ``hot-start'' tracks if the scale-height of the boundary layer is $\gtrsim0.24$, a value not much larger than the scale-height of the circumplanetary disc.  Understanding the
luminosity and radii of young giant planets will thus require detailed
models of circumplanetary accretion.
\end{abstract}





\section{Introduction}
We are now in an era where giant planets orbiting other stars can be
directly imaged. The current sample is small, but in the coming years
the next generation of direct detection instrumentation (e.g. SPHERE -
\citealt{Vigan2010}, GPI - \citealt{Macintosh2008}, HiCIAO -
\citealt{Yamamoto2013}) will grow this sample over the coming
years, and these campaigns recently yielded an actively accreting proto-planet \citep{Sallum2015}. Since sub-stellar mass objects do not generate any internal
luminosity from nuclear burning, they passively cool over time, such
that there is a large degeneracy between luminosity, age and
mass. This issue is particularly acute at young ages when the planets
may not have cooled significantly from their formation conditions, so
that even for a fixed age and mass a planet could admit a large range
of luminosities, depending on its initial thermal content
\citep[e.g.][]{Spiegel2012}.

One can turn this issue on its head and with dynamical constraints on
a planet's mass learn about its initial thermal content and perhaps
its formation. The original approach assumed that giant planets
started cooling from an arbitrary high entropy state
\citep{Stevenson1982,Burrows1997}, with such ``hot start'' models now
typically associated with the outcome of fragmentation in the
protoplanetary disc \citep{Boss1997}. By contrast, \citet{Marley07}
used the standard core-accretion picture
\citep{Pollack1996,Bodenheimer2000} and argued that newly formed
planets would be significantly cooler than the ``hot-start'' models,
with initial cooling times typically $\gtrsim 10^8$~years in these
``cold-start'' models. Observationally, the luminosity of a handful
directly imaged exoplanets with mass constraints are inconsistent with
the ``cold-start'' scenario \citep{Marleau2014} in possible tension
with the core accretion model. On the other hand, it appears difficult
to form  giant planets ($\lesssim 10$ M$_J$) through gravitational
instability \citep{Rafikov2005,Kratter2010}.

While there have been attempts to blur these formation channels into
``warm-start'' scenarios
\citep[e.g.][]{Spiegel2012,Mordasini2012,Mordasini2013}, all such
models implicitly assume that accretion takes place in a spherically
symmetric manner. Accreting material would then be processed by a
shock at the planet's surface, which plays a key role in determining
the initial entropy of the planet \citep[e.g.][]{Marley07}. However,
this is unlikely to be how giant planets accrete their mass in a
realistic scenario. Excess angular momentum of the accreting material
is likely to form a circumplanetary disc through which material can
accrete and be thermally processed. In this work, we argue that disc
accretion can drive the entropies of forming giant planets up to
traditional ``hot-start'' values, even in the core accretion
framework, and that this indeed likely to happen.

\section{Overview of Accretion Paradigm} 
\label{sec:accretion_para}

Most current models of planet formation within the core accretion
scenario assume that the planet remains embedded in the disc until the
disc disperses at some later time. However, it is well known that once
a planet grows to become massive enough to perturb the disc then it
can open a gap \citep[e.g.][]{Lin1993}. A rough estimate of the mass
at which the planet can open a gap is given by the ``thermal-mass'',
when the planet's Hill sphere - $R_H=a(M_p/3M_*)^{1/3}$ - exceeds the
local scale height of the disc ($H$).  For a typical passively heated
protoplanetary discs with temperature $T\propto R^{-1/2}$ \citep{Kenyon1987},
this qualitatively requires:
\begin{equation}
M_p\gtrsim 0.07\mbox{~M}_J~\left(\frac{a}{1\mbox{~AU}}\right)^{3/4}\left(\frac{T_{1{\rm AU}}}{200~\mbox{K}}\right)^{3/2}\left(\frac{M_*}{1\mbox{~M}_\odot}\right)^{-1/2}.
\end{equation}
Therefore, it is likely that a giant planet in the region 1-30~AU will
accrete the majority of its mass after gap opening. Once the
gap has opened, simulations suggest that the incoming accretion
streams possess enough angular momentum to form a circumplanetary disc
\citep[e.g.][]{Dangelo2002,Ayliffe2009,Martin2011,Szulagyi2014}. Thus, the entropy
of the accreting planetary material is no longer associated with the
parent protoplanetary disc, but is rather processed by the
circumplanetary disc and is associated with the accretion process that
transfers material from the disc to the planet. Two such mechanisms
exist: magnetospheric and boundary-layer accretion. Gas giant planets
are hypothesised to have magnetic fields, and indeed Jupiter has a
field strength of $\sim$5~Gauss. In order to determine whether
accretion is controlled by magnetic fields (in the magneto-spherical
accretion model) or whether the disc extends all the way to the
planet's surface we must determine the magnetospheric truncation
radius.

For a formation time $t_{\rm form}=M_p/\dot{M}$, which we expect to be
$\lesssim 10$~Myr, the truncation radius is
\citep{Koenigl1991,Livio1992}:
\begin{eqnarray}
\left(\frac{R_t}{R_p}\right)&\approx& 0.23\left(\frac{B_p}{5\mbox{~Gauss}}\right)^{4/7}\left(\frac{R_p}{10^{10}\mbox{~cm}}\right)^{5/7}\nonumber \\&\times&\left(\frac{t_{\rm form}}{1\mbox{~Myr}}\right)^{2/7}\left(\frac{M_p}{1\mbox{~M}_J}\right)^{-3/7}
\end{eqnarray}
For field strengths similar to gas giants in our solar system this
indicates that the magnetic field cannot truncate the disc and
accretion will proceed all the way to the star. Alternatively, one
would need a planetary field strength in excess of $B_p\gtrsim
65$~gauss for the truncation radius to exceed the planetary
radius \citep[see also][]{Quillen1998,Fendt2003,Lovelace2011,Zhu2015}. The expected magnetic field strengths of young massive giant planets is thought to be $\sim5-12$ times Jupiter's field strength (\citealt{Christensen2009}). Thus, it is highly likely that the circumplanetary disc will accrete onto the
planet through a boundary layer \citep[e.g.][]{LyndenBell1974}.

Since the gas disc lifetime limits the accretion timescale for gas
giants to $\lesssim 10$~Myr we know the accretion rates onto the
protoplanets must be high with values of order $\sim
10^{-9}-10^{-8}$~M$_\odot$~yr$^{-1}$ expected. These accretion rates
can lead to extremely large accretion luminosities
\citep[e.g.][]{Rafikov08b,Owen2014,Zhu2015} of order
$\sim10^{-3}$~L$_\odot$. Approximately, half of this is released in the
disc and the remaining fraction in the boundary layer. Such accretion
luminosities are many orders of magnitude larger than the internal
luminosities of planets in ``cold-start'' scenarios ($\lesssim
10^{-5}$~L$_\odot$) and higher than the majority of ``hot-start''
models \citep[$\lesssim10^{-3}$~L$_\odot$; e.g.][]{Marley07}.  The
ratio of accretion luminosity to internal cooling luminosity is:
\begin{eqnarray}
\frac{GM_p\dot{M}}{F_0 R_p^3}&=& 270 \left(\frac{M_p}{1~{\rm M}_J}\right)\left(\frac{\dot{M}}{10^{-8}~{\rm M_\odot~yr^{-1}}}\right)\nonumber \\ &\times&\left(\frac{L_0}{10^{-4}~{\rm L}_\odot}\right)^{-1}\left(\frac{R_p}{10^{10}~{\rm cm}}\right)^{-1}
\end{eqnarray}
where $F_0$ \& $L_0$ are the internal flux and luminosity of a
passively cooling planet with mass $M_p$ and radius $R_p$.  Therefore,
for a planet forming through disc accretion, the disc and boundary
layer will strongly irradiate the surface of the planet, not unlike
the earliest stages of star formation
\citep[e.g.][]{Adams1986,Rafikov08a}. External irradiation of a gas
giant with an internal luminosity many orders of magnitude smaller is
also similar to the hot Jupiter problem. The irradiation pushes the
radiative-convective boundary deeper into the planetary interior,
preventing the radiation from escaping as easily, thus suppressing
cooling and contraction
\citep{Guillot1996,Burrows2000,Arras2006}. Therefore, by accreting
through a disc, the planet cooling will be suppressed and it will
retain a higher entropy than if it were cooling passively.

\section{Simple Disc-Fed Planetary Formation Model}

We assume that the bulk of the planet's mass is contained in a
convective envelope surrounding the core. We can evaluate the binding
energy of this envelope by assuming that the envelope mass exceeds the
core mass and that its structure is described by a polytrope with
$n=3/2$, so that the total binding energy ($E_p$) is:
\begin{equation}
E_p=-\frac{3}{7}\frac{GM_p^2}{R_p},
\end{equation}
where $M_p$ \& $R_p$ are the planet's mass and radius,
respectively. As the planet accretes, its total binding energy
evolves. Following \citet{Hartmann97}, who studied the evolution of
accreting low-mass stars, we assume the bulk of the planet remains
convective and describe the evolution as:
\begin{equation}
\frac{{\rm d}E_p}{{\rm d}t}=(\epsilon-1)\frac{GM_p\dot{M}}{R_p}-L_{\rm rad}.\label{eqn:Eevolve}
\end{equation}
The accretion efficiency parameter, $\epsilon$, represents the
internal energy of the accreted matter, $\dot{M}$ is the accretion
rate and $L_{\rm rad}$ describes the radiative losses from the
planetary surface. The boundary layer can be described using a
``slim-disc'' model \citep[e.g.][]{Abramowicz88,Popham91}. Global
integration of the slim-disc energy equation \citep{Popham97} shows
that\footnote{Assuming the discs luminosity is large compared to
  $(H/R)L_p$}:
\begin{equation}
L_d+C_p\dot{M}c_s^2=\frac{GM_p\dot{M}}{R_p}\left(1-j\frac{\Omega_p}{\Omega_K}+\frac{1}{2}\frac{\Omega_p^2}{\Omega_K^2}\right)
\end{equation}
where $L_d$ is the luminosity radiated away by the disc surface
layers, $C_p$ is the heat capacity, $\Omega_p$ is the angular velocity
of the planet, $\Omega_K$ is the Keplerian angular velocity at the
radius of the planet and $j$ is the angular momentum flux normalised
to $\dot{M}\Omega_KR_p^2$. {\bc Here we do not attempt to model the evolution of the planetary rotation rate, assuming it is not close to break-up, but note that given an explicit boundary layer model the evolution of the planet’s angular velocity could be computed as well.} The second term of the left-hand side
represents the  energy advected into the planet and the right hand
side represents the energy dissipated by the disc and boundary
layer, {\bc where we neglect the role of the $\Omega_p/\Omega_k$ term}. Therefore, the fraction $\epsilon$ of the accretion luminosity
advected into the star is \citep{Popham97}:
\begin{equation}
\epsilon=C_p\dot{M}c_s^2\left(\frac{GM_p\dot{M}}{R_p}\right)^{-1}=C_p\left(\frac{H_p}{R_p}\right)^2
\end{equation}
where $H_p$ is the scale height of the boundary layer at the planetary
radius.

\begin{figure}
\centering
\includegraphics[width=\columnwidth]{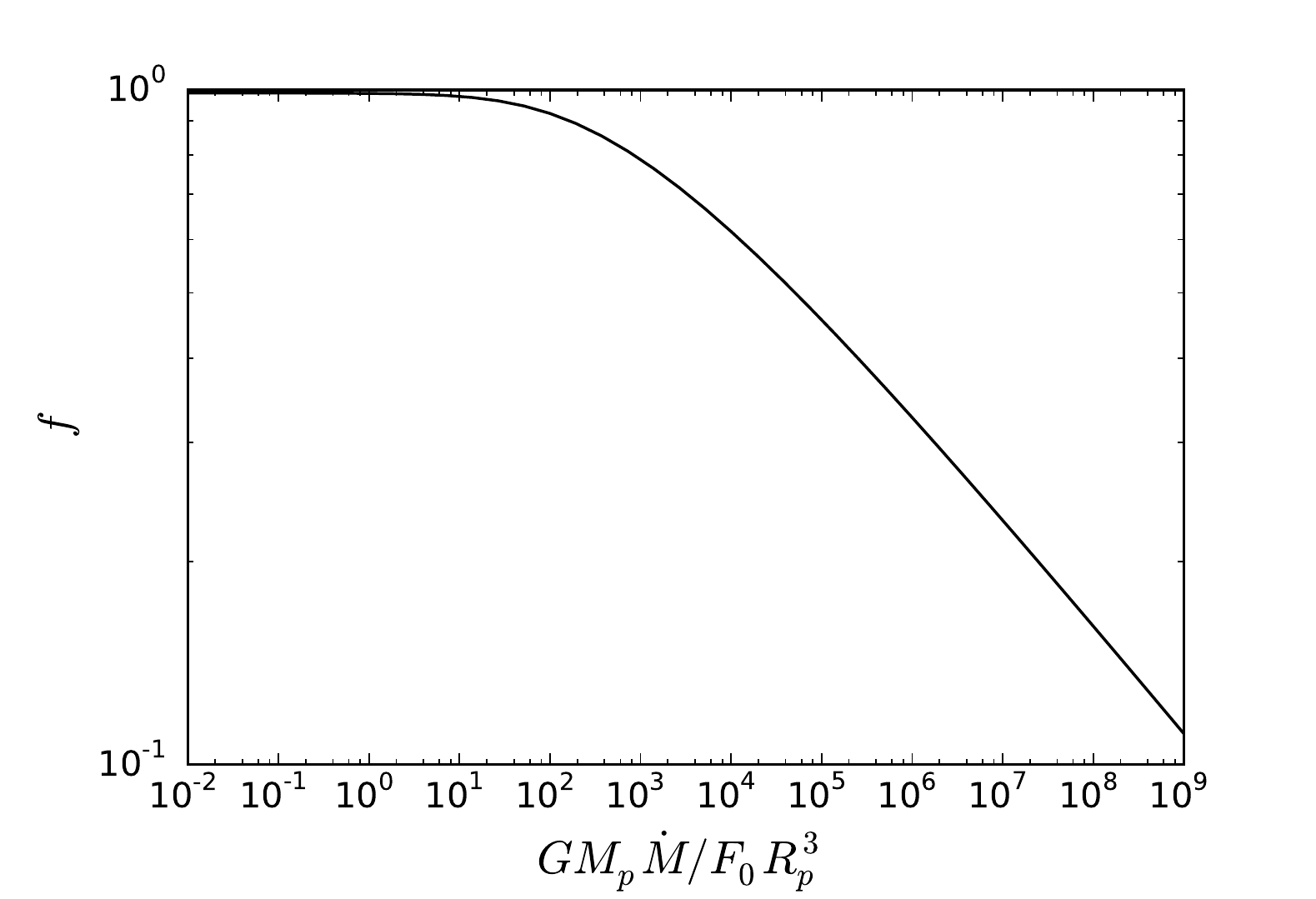}
\caption{The luminosity suppression factor $f$ as a function of $GM_p\dot{M}/F_0 R_p^3$. For accreting planets, this value is typically in the range $1-10^4$.}\label{fig:f_lambda}
\end{figure}

Finally, assuming that the boundary layer obscures an area $4\pi R_pH_p$ of the planetary surface, $L_{\rm rad}$ is given by:
\begin{equation}
L_{\rm rad}= 4\pi R_p^2\left(1-\frac{H_p}{R_p}\right)\int_0^{\pi/2}\!\!\!\! F_p(\theta){\rm d}\cos\theta \label{eqn:Lrad1}
\end{equation}
where $F_p(\theta)$ is the flux emerging from the planetary surface at
an angle $\theta$ to the pole. As discussed above, the accretion
luminosity often exceeds the internal luminosity by several orders of
magnitude. \citet{Rafikov08a} showed that the integral in
Equation~\ref{eqn:Lrad1} can be written as:
\begin{equation}
\int_0^{\pi/2}\!\!\!\! F_p(\theta){\rm d}\cos\theta=F_0\,f\left(\frac{G M_p \dot{M}}{F_0 R_p^3}\right)
\end{equation}
Calculation of $F_p(\theta)$ can be performed assuming  emission
from a standard $\alpha$-disc
\citep[e.g.][]{Adams1986,Popham97,Rafikov08a}. The function $f$ is not
analytic; however, given an opacity law of the form $\kappa\propto P^a
T^b$, $f$ can be calculated following \citet{Rafikov08a}. A power-law
fit to the opacity law in the planetary regime is given by
\citet{Rogers10} with $a=0.68$ \& $b=0.45$. Adopting this fit, the
function $f$ is shown in Figure~\ref{fig:f_lambda}, indicating that
irradiation from the disc can lead to cooling luminosities several
tens of percent lower than for a passively cooling planet.
Therefore, we write $L_{\rm rad}$ as:
\begin{equation}
L_{\rm rad}=fL_0\left(1-\frac{H_p}{R_p}\right),\label{eqn:Lrad2}
\end{equation}
where $L_0$ is the luminosity of an isolated planet. 

\begin{figure*}
\centering
\includegraphics[width=\textwidth]{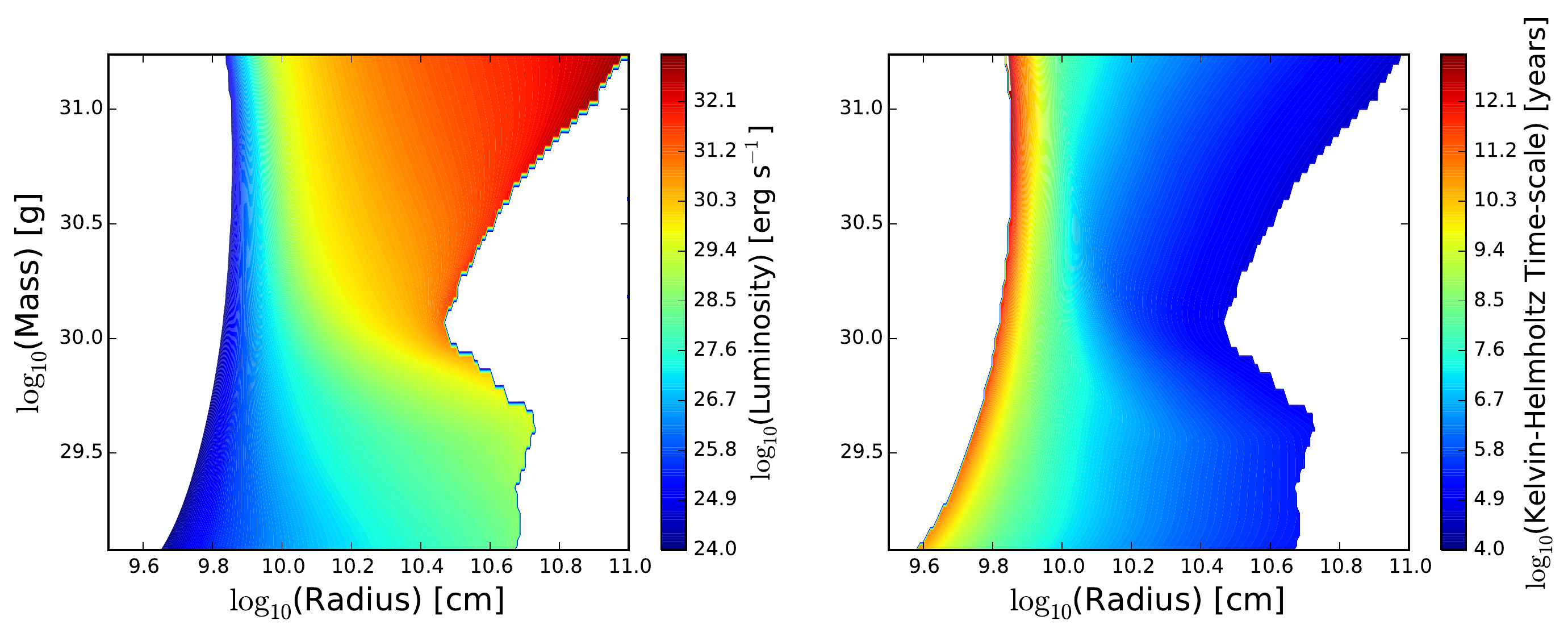}
\caption{The luminosity (left) and Kelvin-Helmholtz time-scale (right) as a function of planet mass and radius. White regions are areas where no models were calculated.}\label{fig:mesa_results}
\end{figure*}

In order to fully evolve our accreting planet-disc system we need to
know the scale height of the boundary layer at the planetary
surface. Boundary layers are poorly understood; since ${\rm
  d}\Omega^2/{\rm d}R>0$ they are stable to the MRI and an
$\alpha$-viscosity prescription may lead to unphysical solutions
\citep{Pringle1977,Popham1992}. Recently
\citet{Belyaev2013a,Belyaev2013b} showed that angular momentum
transport can occur through waves arising from the sonic instability
in boundary layers; however, a simple boundary layer model including
this mechanism does not exist. We can still make some reasonable
estimate of the scale height by using the disc scale height just
outside the boundary layer. Since one would expect enhanced
dissipation in the boundary layer due to the sharper angular momentum
gradients present, we might expect the scale height to be larger in
the boundary layer than in the disc. For an actively heated disc, {\bc where one equates radiative cooling with local viscous dissipation in a Keplerain disc}, the
scale height in the disc, $H^d$, is given by:
\begin{eqnarray}
\frac{H^d_p}{R_p}&\approx&0.15~\mu^{-1/2}\left(\frac{M_p}{1~{\rm M_J}}\right)^{-3/8}\left(\frac{\dot{M}}{10^{-8}~{\rm M_\odot~yr^{-1}}}\right)^{1/8}\nonumber\\&\times&\left(\frac{R_p}{10^{10}~{\rm cm}}\right)^{1/8},\label{eqn:Hp}
\end{eqnarray}
where $\mu$ is the mean particle weight in amu. {\bc In obtaining Equation~\ref{eqn:Hp} we have neglected the ``standard'' $(1-\sqrt{R_p/R})$ factor. This factor arises if one applies a zero-torque boundary condition at the boundary-layer/disc interface, resulting in a decrease in the surface density towards the planet's surface. This specific boundary condition is unlikely to be appropriate for the boundary-layer/disc interface which is more likely to have an angular velocity and thus associated torque of a full Keplerian disc \citep[e.g.][]{Popham91}. } Since Equation~\ref{eqn:Hp} is weakly sensitive to the input
parameters, we assume that the scale height of the boundary
layer\footnote{Since we only evaluated the scale height in the disc
  not the boundary layer we cannot verify this assumption.} remains
constant for the entire evolution and choose values $\gtrsim0.15$. Combining Equations~\ref{eqn:Eevolve} \& \ref{eqn:Lrad2}, we
obtain the evolution equation for the radius of an accreting
convective planet:
\begin{equation}
\frac{\dot{R}_p}{R_p}=\frac{7}{3}\frac{\dot{M}}{M_p}\left[C_p\left(\frac{H_p}{R_p}\right)^2-\frac{1}{7}\right]-\frac{7}{3}\frac{L_{\rm rad}R_p}{GM_p^2}\label{eqn:evolve}
\end{equation}
To integrate Equation~\ref{eqn:evolve}, one needs to know the passive
luminosity of the planet as a function of planet mass and radius. To
calculate this we use the {\sc mesa} code
\citep{Paxton2011} to produce a series of hydrostatic
models as a function of planet mass and radius, assuming a
10~M$_\oplus$ rocky core and the \citet{Freedman2008} opacities. The
resulting luminosities (left-panel) and Kelvin-Helmoltz time-scales
($t_{\rm KH}$, right-panel) are shown in
Figure~\ref{fig:mesa_results}.

We can now understand how the proto-planet will evolve from
Equation~\ref{eqn:evolve}. The first term on the RHS is $\sim
1/t_{\dot{M}}$, where $t_{\dot{M}}$ is the mass evolution time-scale,
while the second term is approximately $1/t_{\rm KH}$. Therefore, if
$t_{\rm KH}<t_{\dot{M}}$ the proto-planet will cool and shrink. If
$t_{\dot{M}}<t_{\rm KH}$ then the evolution of the planet depends on
the internal energy of the incoming material, and as such there is a critical
$H_p/R_p$ above which the planet increases in radius as it accretes,
otherwise it shrinks. This critical boundary layer height is
$H_p/R_p=\sqrt{(\gamma-1)/7\gamma}$, which is $\approx0.24$ for
$\gamma=5/3$ and $\approx0.2$ for $\gamma=7/4$. Given typical
temperatures in the boundary layer, we expect the gas to be monotonic,
so that for $H_p/R_p\gtrsim0.24$ the proto-planet's radius will
increase if $t_{\dot{M}}<t_{\rm KH}$. This critical value of the
boundary layer scale height is not much larger than the estimated disc
scale height (Equation~\ref{eqn:Hp}), thus we may expect the planet's
radius to increase rather than decrease during circumplanetary disc
accretion. The left panel of Figure~\ref{fig:mesa_results} shows that,
in the giant planet regime $M_p\gtrsim10^{30}$~g, luminosity
increases with increasing radius as the mass increases. Therefore,
disc accretion through a boundary layer can potentially drive the
planet to high luminosities, comparable or even larger than typical
values in the ``hot-start'' scenarios.

\section{Results} 

We consider the evolution of a proto-planet accreting through a
boundary layer. The planet is taken to be an initially 50~M$_\oplus$
planet with a 10~M$_\oplus$ core and we assume the core doesn't grow
in mass during the evolution. We assume that the accretion rate is
constant and the radius of the planet evolves according to
Equation~\ref{eqn:evolve}, where the relative scale height of the
boundary layer to the planet's radius remains fixed for the entire
accreting period. We consider a variety of formation times, final
masses, initial cooling times for the 50~M$_\oplus$ proto-planet and
$H_p/R_p$ values.

\begin{figure*}
\centering
\includegraphics[width=\textwidth]{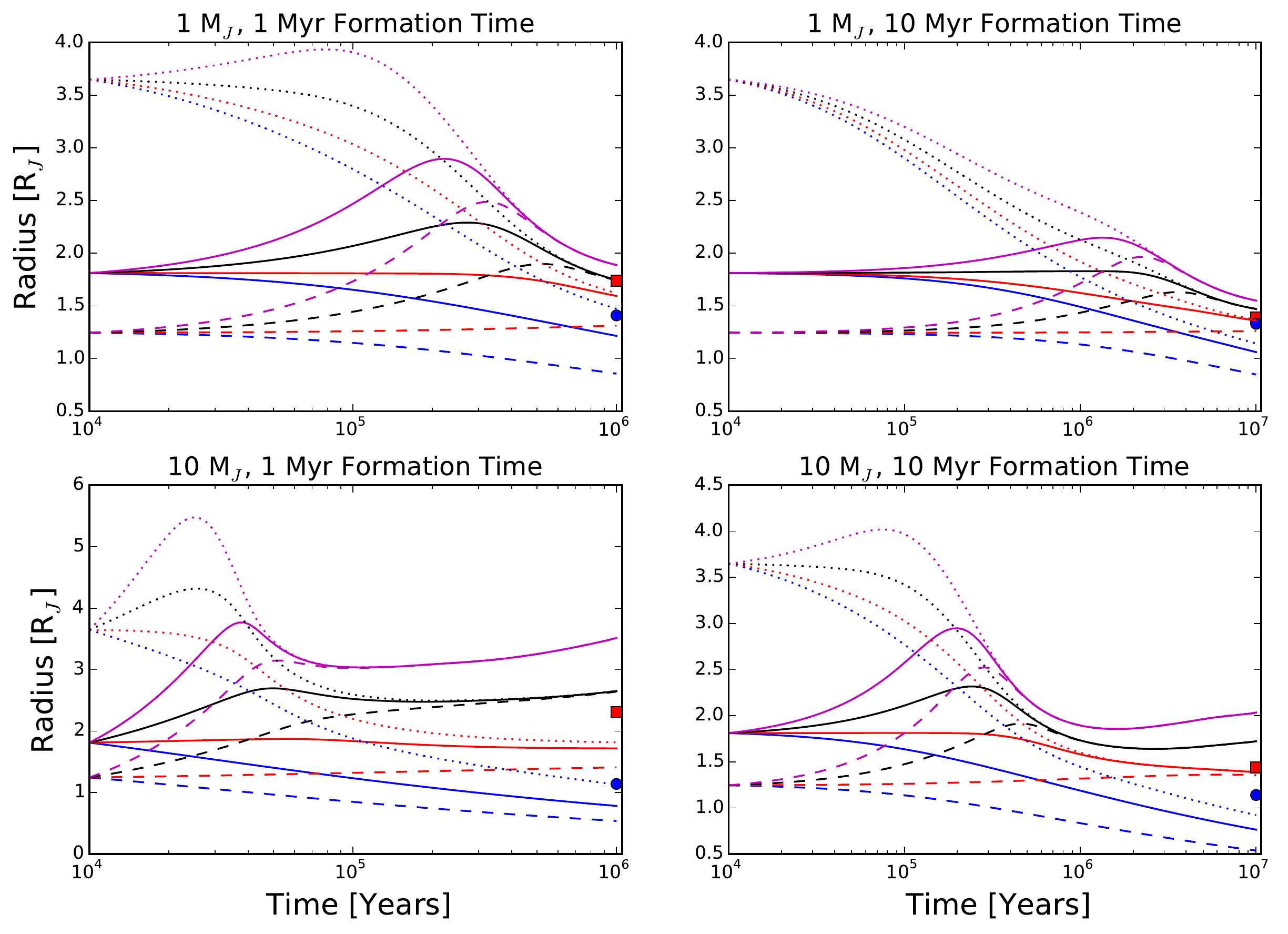}
\caption{The radius evolution of an accreting proto-planet with a
  final mass of 1~M$_J$ (top) and 10~M$_J$ (bottom) accreting at fixed
  $\dot{M}$ over a time of 1~Myr (left) and 10~Myr (right). The
  initial cooling times of the proto-planets are $10^6$ (dotted), $10^7$
  (solid) and $10^8$~years (dashed). The constant values adopted for $H_p/R_p$
  are 0.15 (blue), 0.25 (red), 0.35 (black) \& 0.45
  (magenta). The square and circular points show the \citet{Spiegel2012} ``hot'' and ``cold'' start planet radii respectively, for the given age and final planet mass.}\label{fig:1_10Mj}
\end{figure*}

The radius evolution of a proto-planet with final mass 1~M$_J$ (top)
and 10~M$_J$ (bottom), accreting at a constant rate with a formation
time of 1~Myr (left) and 10~Myr (right), are shown in
Figure~\ref{fig:1_10Mj}. Three initial cooling times of $10^6$~years,
$10^7$~years \& $10^8$~years are shown each with $H_p/R_p$ values of
0.15, 0.25, 0.35, 0.45.

The accretion of material by a proto-planet via a circumplanetary disc
and boundary layer can result in a large range of initial planet
properties, from ``hot start'' to ``cold start'' conditions. For
reference,  ``hot start'' and the
``cold start'' radii from \citet{Spiegel2012} are shown as the points. In all cases $t_{\rm KH}>t_{\dot{M}}$ initially and the proto-planet follows an evolutionary
path determined by $H_p/R_p$. Therefore, above the critical value of
$H_p/R_p$ the proto-planet grows in radius and below it shrinks in
radius, as expected. For the highest $H_p/R_p$ values considered, the
proto-planet is driven to high enough luminosities that radiative
cooling dominates and contraction ensues. This decrease in radius
causes the planets to follow a convergent evolutionary path and the
final properties of planets with $H_p/R_p\gtrsim0.3$ are insensitive
to their initial properties. However proto-planets forming with
$H_p/R_p\lesssim0.2$ do not cool appreciably and final properties
post-accretion bear the signature of the initial Kelvin-Helmholtz
time-scales. For massive planets (10~M$_J$ final masses), the radius
starts to increase again once they reach about 2~$M_J$. This can be
understood from Figure~\ref{fig:mesa_results}, which shows that, at
high masses, the cooling time increases as one increases mass at fixed
radius. So a planet whose evolution was dominated by cooling at lower
masses can evolve into a region where cooling becomes subdominant and
the thermal content of the accreted material drives the evolution
instead. Unsurprisingly, faster accreting proto-planets experience
stronger evolution, as mass accretion dominates over cooling more
prominently.

\begin{figure*}
\centering
\includegraphics[width=\textwidth]{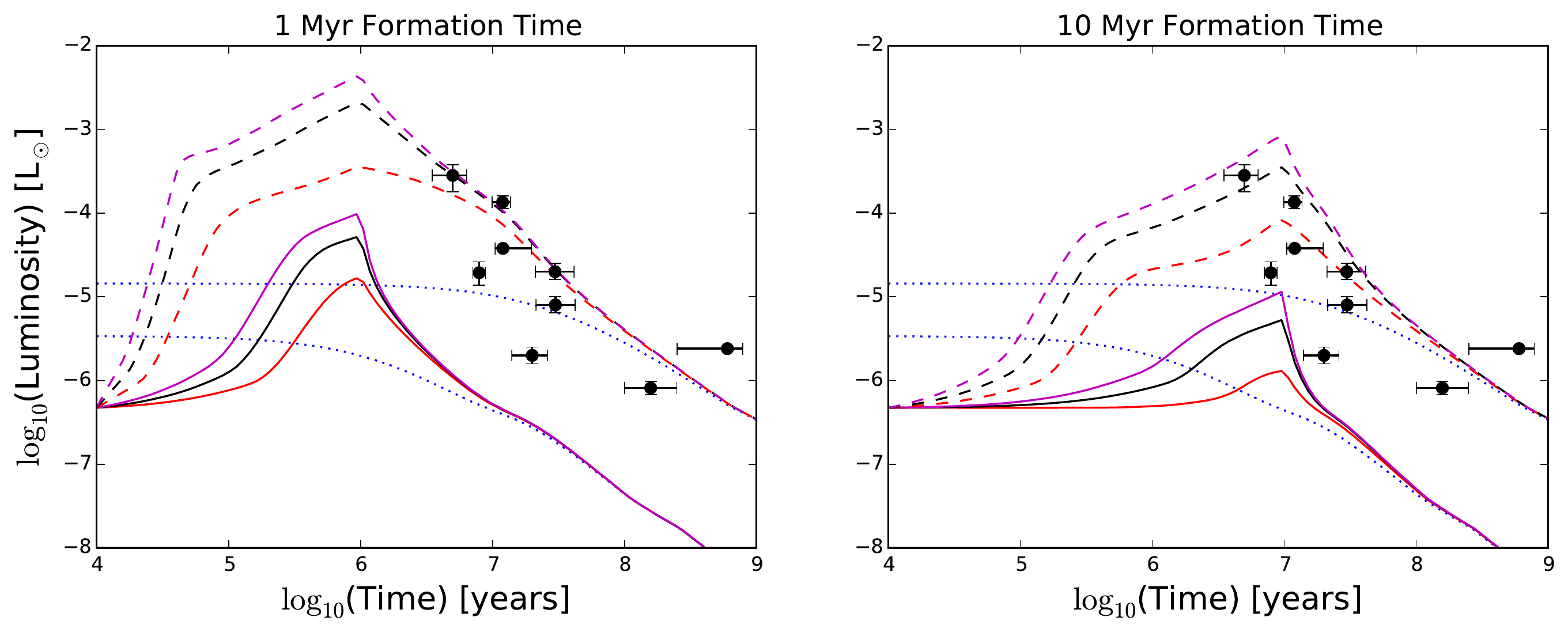}
\caption{Luminosity evolution of young planets through formation and
  subsequent cooling. The solid lines show a planet that reaches
  1~M$_J$ and the dashed lines a planet that reaches 10~M$_J$ at the
  end of the accretion phase which lasts 1~Myr (left) or 10~Myr
  (right). The constant values adopted for $H_p/R_p$ during the
  accretion phase are 0.25 (red), 0.35 (black) \& 0.45 (magenta). The
  blue dotted lines are ``cold-start'' models which have passively
  evolve from an initial cooling time of $10^8$~years. The points
  represent observed exoplanets (objects consistent with $<25$~M$_J$
  tracks) taken from the sample compiled by
  \citep{Neuhauser2012,Bowler2013}.}\label{fig:lum_compare}
\end{figure*}

Finally, we can compute the luminosity evolution of our planets,
including at late times after accretion stopped.  We do this by
continuing to evolve the planet according to
Equation~\ref{eqn:evolve}, however, we set $\dot{M}=0$ and allow the
planet to cool over its entire surface. The comparison of model tracks
with the handful of directly imaged giant planets for which their
luminosity and age has been measured is shown in
Figure~\ref{fig:lum_compare}. Planetary formation through boundary
layer accretion can drive planets onto ``hot-start''-like evolutionary
paths , which can successfully match the luminosity of young directly
imaged exoplanets.

\section{Discussion and Summary}
A proto-planet embedded in a proto-planetary disc is likely to open a gap at modest
masses, thus a giant planet is likely to accrete the majority of its
mass through a circumplanetary disc. This scenario is unlike the
direct collapse (gravitational fragmentation) framework where the high
entropy disc material is turned into a planetary mass object on a
short-timescale, in an essentially adiabatic fashion. It is also
unlike the final stages of the standard core accretion framework,
which assumes that the proto-planet accretes spherically, shocks and
then radiate away the disc material's entropy, accreting material with
the same temperature as the proto-planet's atmosphere, leading to
comparatively small, low luminosity planets
\citep[e.g.][]{Marley07}. In the disc accretion scenario considered
here, we argue that the entropy of the accreting material depends on
how the material is transported from the disc to the planet. Unless
the proto-planet has a very strong magnetic fields ($\gtrsim
65$~gauss) this will occur through a boundary layer and the fraction
of heat added to the proto-planet directly depends on the thermal
structure of the boundary layer, with puffier boundary layers
advecting more heat into the planet. There is a critical value of the
height of the boundary layer above which the proto-planet becomes
inflated by accretion and driven to high luminosities, $H_p/R_p\gtrsim
0.24$, which is only slightly larger than the scale height of the
circumplanetary disc feeding the boundary layer. To the extent that
boundary layers are hotter than their circumplanetary disc, one may
expect in the majority of cases that circumplanetary disc accretion
will inflate the proto-planet, driving it to high luminosities in line
with those expected from ``hot-start'' or direct collapse scenarios.

Before concluding, we note that our models of disc-fed planet
formation are highly idealized. We have assumed an $n=3/2$
polytrope. For the largest boundary layers we are adding extremely
high entropy gas on-top of lower entropy gas, so that the envelope’s
outer layers may become stably stratified. Accretion over a long
period of time could result in a complicated layered structure of
convective and radiative zones, meaning that an $n=3/2$ polytrope may
no longer be a good description of the proto-planetary structure. {\bc If such a stably stratified region were very thin it may cool quickly, resulting in low luminosity planets fairly soon after accretion ceases. Simple models -- a constant opacity, hydrostatic radiative envelope \citep[c.f.][]{Stevenson1982} --  suggest a thick radiative layer with large thermal inertia, so that the inflated radii should remain that way for a long time after accretion stops.  More detailed models are needed to address this reliably.} One
obvious avenue for progress is thus to couple boundary-layer accretion
models with detailed structure calculations for the planetary
interior.

In conclusion, disc-fed accretion may be the dominant mechanism
shaping the properties of young giant planets, where the thermal
structure of the boundary layer controls the amount of heat advected
into the planet during late-stage growth. Such a process will
naturally arise in the core-accretion scenario once the planet becomes
massive enough to open a gap in the protoplanetary disc and angular
momentum conservation leads to the formation of a disc around the
proto-planet. It therefore seems that further development of boundary
layer accretion theory is warranted to understand giant planet
formation and to interpret the wealth of direct imaging data expected
to become available over the next few years.



\acknowledgements
\paperacknowledge

\bibliographystyle{apj}

\end{document}